\documentclass[preprint, superscriptaddress,amsmath, nofootinbib]{revtex4-1}
\pdfoutput=1
\usepackage{dcolumn}
\usepackage{bm}
\usepackage{graphicx}
\usepackage{amssymb,amsmath}
\usepackage{multirow}
\usepackage{units,changes}
\usepackage{color,url}
\usepackage{tabu}
\usepackage{array}
\usepackage[colorlinks=true,urlcolor=blue,anchorcolor=blue
,citecolor=blue,filecolor=blue,linkcolor=blue,menucolor=blue
,linktocpage=true,pdfproducer=medialab,pdfa=true]{hyperref}
\usepackage{colordvi}
\usepackage{subcaption}
\def\beqa{\begin{eqnarray}}
\def\eeqa{\end{eqnarray}}

\usepackage{soul}

\begin{document}
\preprint{CPTNP-2025-038}

\title{Searching for axion-like particles from tau exotic decays at the Super Tau-Charm Facility and its far detectors}
\def\slash#1{#1\!\!\!/}


\author{Xu-Hui Jiang}
\email{jiangxh@ihep.ac.cn}
\affiliation{Center for Future High Energy Physics, Institute of High Energy Physics,
Chinese Academy of Sciences, Beijing 100049, China}
\affiliation{China Center of Advanced Science and Technology, Beijing 100190, China}

\author{Chih-Ting Lu}
\email{ctlu@njnu.edu.cn}
\affiliation{Department of Physics and Institute of Theoretical Physics, Nanjing Normal University, Nanjing, 210023, P. R. China}
\affiliation{Nanjing Key Laboratory of Particle Physics and Astrophysics, Nanjing, 210023, China}


\begin{abstract}
Leptophilic axion-like particles (ALPs) extend the Standard Model (SM) with brand-new interactions between the ALP and leptons. In this work, we focus on studying exotic $\tau$ decays to explore such ALPs. Both tauphilic and lepton flavor universal (LFU) scenarios, with electroweak preserving and violating benchmarks, have been investigated at a future $\tau$-factory, namely the Super Tau-Charm Facility (STCF) under development in China. Both main and far detection are proposed, targeting leptonic decays $\tau$, $\tau^- \to \ell^-\bar\nu_\ell \nu_\tau a$. For main detection, the $\tau^+\tau^-$ threshold $\sqrt s=3.56$ GeV and the higher central energy $\sqrt s=4.2$ GeV are taken into account. In addition, a cylinder-like far detector has been proposed to complement main detection. We demonstrate the huge potential to detect leptophilic ALPs at the STCF. Concretely, for the tauphilic scenario, the STCF is mostly sensitive in the currently unprobed region with $m_a\sim 1000$ MeV. However, in the LFU scenario, the dilepton channel tremendously shortens the ALP lifetime, and eventually the STCF only allows a precise measurement in the new regime with $100~\text{MeV}\lesssim m_a\lesssim 200~\text{MeV}$.
\end{abstract}

\maketitle




\newpage 
\section{Introduction}

Axion-like particles (ALPs) are pseudoscalar fields that arise as Nambu–Goldstone bosons associated with a spontaneously broken global $U(1)$ symmetry, characterized by a new physics scale $\Lambda$~\cite{Graham:2015ouw,Choi:2020rgn}. Theoretically, ALPs are predicted in a variety of beyond the Standard Model (SM) scenarios, including extra-dimensional constructions~\cite{Chang:1999si,Bastero-Gil:2021oky}, string compactifications~\cite{Svrcek:2006yi,Arvanitaki:2009fg,Cicoli:2012sz,Visinelli:2018utg}, dark quantum chromodynamics (QCD) models through hadronization~\cite{Cheng:2021kjg,Cheng:2024hvq,Cheng:2024aco}, and other ultraviolet completions~\cite{Bagger:1994hh,Yamada:2021uze,Choi:2024ome}. Phenomenologically, ALPs offer compelling solutions to several open questions in particle physics and cosmology. The QCD axion, a well-motivated ALP candidate, provides an elegant explanation for the strong CP problem~\cite{Peccei:1977hh,Peccei:1977ur,Weinberg:1977ma,Wilczek:1977pj,Kim:1979if,Shifman:1979if,Dine:1981rt}. Ultralight ALPs are also viable dark matter candidates~\cite{Preskill:1982cy,Abbott:1982af,Dine:1982ah}. Furthermore, ALPs appear in models addressing the hierarchy problem of Higgs mass~\cite{Graham:2015cka}, baryogenesis~\cite{Jeong:2018jqe,Domcke:2020kcp,Im:2021xoy,Co:2024oek}, and certain experimental anomalies~\cite{Chang:2021myh,Cheung:2024kml,Co:2024oek}. Thus, ALPs represent a well-founded portal to new physics, and their discovery could shed light on high-energy scales currently inaccessible by direct exploration. However, their low masses and feeble interactions with SM particles make the experimental detection of ALPs particularly challenging.

The ALP mass $m_a$ is generally assumed to lie far below the symmetry-breaking scale $\Lambda$, though its possible values span a wide range, from effectively massless up to the electroweak scale and beyond. In the framework of effective field theory (EFT)~\cite{Brivio:2017ije,Bauer:2017ris,Bauer:2018uxu,Ebadi:2019gij,Bauer:2020jbp,Bauer:2021mvw}, the couplings of ALPs to SM fields can be treated independently, enabling diverse experimental search strategies tailored to different mass regimes. For very light ALPs ($0 < m_a \lesssim $ MeV), experiments such as light-shining-through-walls (LSW)~\cite{Sikivie:1983ip,Beyer:2021mzq} and astrophysical observations~\cite{Raffelt:1990yz,Irastorza:2011gs,Marsh:2015xka,CAST:2017uph,Kavic:2019cgk, Alda:2024cxn} provide the leading sensitivity. In the keV-GeV range, cosmological probes constrain regions of parameter space with extremely weak couplings~\cite{Marsh:2015xka,OHare:2024nmr}, while fixed-target experiments and low-energy $e^+ e^-$ colliders are sensitive to scenarios with larger interactions~\cite{Riordan:1987aw,Bjorken:1988as,BaBar:2011kau,Dobrich:2015jyk,Dobrich:2019dxc,Belle-II:2020jti,Zhevlakov:2022vio,Niedziela:2024khw}. For heavier ALPs (GeV $\lesssim m_a\lesssim $ TeV), high-energy colliders offer a powerful means of detection~\cite{Brivio:2017ije,Bauer:2017ris,Bauer:2018uxu,Ebadi:2019gij}. Despite existing constraints across broad regions of the ALP parameter space, substantial gaps remain, particularly for specific mass values and coupling structures, motivating the development of new search strategies to probe these uncharted territories. In particular, leptophilic ALPs, which couple predominantly to SM charged leptons, have recently attracted increased attention~\cite{Kirpichnikov:2020lws,Bollig:2020xdr,Croon:2020lrf,Han:2020dwo,Chang:2021myh,Carenza:2021pcm,Cheung:2021mol,Caputo:2021rux,Bertuzzo:2022fcm,Cheung:2022umw,Lucente:2022esm,Altmannshofer:2022ckw,Lu:2022zbe,Calibbi:2022izs,Lu:2023ryd,Buonocore:2023kna,Calibbi:2024rcm,Budhraja:2024diy,Wang:2024zky,Jiang:2024cqj,Li:2025yzb,Li:2025beu,Yue:2025ksr,Huang:2025rmy,Ema:2025bww} compared to the more extensively studied ALP interactions with SM gauge bosons and quarks.

The Super Tau-Charm Facility (STCF) is a next-generation high-intensity electron-positron collider under development in China, designed to operate in the tau-charm energy region with a center-of-mass energy range of 2–7 GeV and a projected peak luminosity exceeding $0.5 \times 10^{35}~\text{cm}^{-2}\text{s}^{-1}$~\cite{Bao:2025aic}. This facility will produce the world’s largest samples of tau leptons and charm hadrons, enabling unprecedented precision measurements in non-perturbative QCD, charm physics, and tau lepton decays~\cite{Achasov:2023gey}. A primary physics motivation for the STCF is to address fundamental questions such as the mechanism of strong interaction confinement, the origin of CP violation, potentially illuminating the matter–antimatter asymmetry of the universe, and searches for physics beyond the Standard Model (BSM), including lepton flavor violation, ALPs, dark photons, and sterile neutrinos~\cite{Shi:2019vus,Zhang:2019wnz,Fan:2021mwp,Liu:2021qio,Xiang:2023mkc,Zeng:2023wqw,Lu:2023cet}.

The copious production of tau leptons at the STCF offers a unique opportunity to investigate exotic decay channels that are presently inaccessible at other facilities. In this work, we focus on ALP–tau interactions through the exotic decay mode $\tau^-\to a\ell^-\bar{\nu_{\ell}}\nu_{\tau}$, which remains relatively unconstrained by existing experiments~\cite{Cui:2021dkr,Ema:2025bww}. Depending on the electroweak symmetry structure of the EFT describing ALPs, two distinct interaction types contribute to this process: in the electroweak-preserving scenario, only the $a\tau^+\tau^-$ coupling is relevant, whereas the electroweak-violating scenario introduces an additional four-point interaction $a–W–\ell–\nu_{\ell}$~\cite{Altmannshofer:2022ckw}. We further consider two representative ALP decay scenarios: a tauphilic scenario, in which the ALP decays predominantly to diphoton final states and may exhibit macroscopic lifetimes, and a lepton flavor universal (LFU) scenario, where ALPs decay mainly to electron or muon pairs, leading to significantly shorter lifetimes. We analyze the sensitivity of displaced ALP decays within the STCF main detector, as well as at a proposed far detector. This study also represents the first proposal for a far detector at the STCF, a setup that could potentially enhance sensitivity to light weakly interacting BSM particles.

The remainder of this paper is organized as follows. In Sec.~\ref{sec:alpproduction}, we review the EFT framework for ALP–lepton interactions and their implications for production and decay dynamics. Section~\ref{sec:tas} presents the sensitivity study for long-lived ALPs from tau exotic decays in both tauphilic and LFU scenarios, considering both the main STCF detector and a far detector setup. We conclude and summarize our results in Sec.~\ref{sec:conclusion}.

\section{ALP-Lepton Interactions: Productions and Decays}\label{sec:alpproduction}

The ALP, denoted as $a$, arises from breaking of the global $U(1)_{\text{PQ}}$ symmetry~\cite{Peccei:1977hh}, which introduces a shift symmetry in the Lagrangian, $a(x)\to a(x) +\text{const}$. The interaction between the ALP and leptons would be expressed as $\partial_{\mu} a J^{\mu}_{\text{PQ},\ell}$, where $J^{\mu}_{\text{PQ},\ell}$ stands for the general lepton current associated with the global $U(1)_{\text{PQ}}$ symmetry, given by~\cite{Bertuzzo:2022fcm,Altmannshofer:2022ckw,Lu:2022zbe, Jiang:2024cqj},
\begin{equation} 
\label{eq:Jint}
J^{\mu}_{\text{PQ},\ell} = \frac{c^V_{\ell}}{2\Lambda}\overline{\ell}\gamma^{\mu}\ell + \frac{c^A_{\ell}}{2\Lambda}\overline{\ell}\gamma^{\mu}\gamma_5\ell + \frac{c_{\nu}}{2\Lambda}\overline{\nu_{\ell}}\gamma^{\mu} P_L \nu_{\ell}~, 
\end{equation} 
where $\Lambda$ represents the scale of $U(1)_{\text{PQ}}$ symmetry breaking, and $c^V_{\ell}$, $c^A_{\ell}$, and $c_{\nu}$ are dimensionless couplings. The symbols $\ell$ and $\nu$ denote charged leptons and neutrinos, respectively. Throughout the study, we take $\Lambda=1$ TeV as a reference. With integration by parts and equations of motion, $\partial_{\mu} a J^{\mu}_{\text{PQ},\ell}$ can be transformed into $a\partial_{\mu}J^{\mu}_{\text{PQ},\ell}$. As a result, the full Lagrangian can be formulated as~\cite{Altmannshofer:2022ckw,Lu:2022zbe, Jiang:2024cqj}:
\begin{align} 
& a ~\partial_{\mu}J^{\mu}_{\text{PQ},\ell} = i c^A_{\ell}\frac{m_{\ell}}{\Lambda}~a\overline{\ell}\gamma_5\ell \label{eq:int} \\
& + \frac{\alpha_{\text{em}}}{4\pi\Lambda} \bigg[  \frac{ c^V_{\ell} -c^A_{\ell} + c_{\nu}}{4 s^2 _W}~a W^{+}_{\mu\nu}\tilde W ^{-,\mu\nu} \notag \\ 
& + \frac{c^V_{\ell} - c^A_{\ell} (1 -4 s^2_W)}{2s _W c_W}~a F_{\mu\nu}\tilde{Z} ^{\mu\nu} - c^A_{\ell}~a F_{\mu\nu} \tilde{F}^{\mu\nu} + \notag \\ 
& \frac{c^V_{\ell} (1 -4 s^2_W) -c^A_{\ell} (1 -4 s^2_W +8 s^4_W)  + c_{\nu}}{8 s^2_W c^2_W}~a Z_{\mu\nu}\tilde{Z}^{\mu\nu}\bigg]  \notag \\ 
& - \frac{ig}{2\sqrt{2}\Lambda}(c^A_{\ell} - c^V_{\ell} + c_{\nu})~a (\bar\ell \gamma^{\mu} P _L \nu) W_{\mu}^{-} ~+~\text{h.c.}  \,. \notag 
\end{align}
Here, $m_{\ell}$ is the charged lepton mass, $\alpha_{\text{em}}$ is the fine structure constant, $s_W$ and $c_W$ are the sine and cosine of the weak mixing angle, respectively, and $g$ is the weak coupling constant. The symbols $W^{\pm}_{\mu\nu}$, $F_{\mu\nu}$, and $Z_{\mu\nu}$ are the field strength tensors for $W^{\pm}$, $\gamma$, and $Z$, respectively. The dual field strength tensor is defined as $\tilde{F}_{\mu\nu}=\frac{1}{2}\epsilon_{\mu\nu\rho\sigma}F^{\rho\sigma}$. Details of the derivation are found in the Appendix of Ref.~\cite{Jiang:2024cqj}. 

According to the ALP-lepton interactions in Eq.~(\ref{eq:int}), the ALP can decay directly to a pair of leptons or a pair of gauge bosons due to the chiral anomaly and one-loop triangle Feynman diagrams. For ALPs with masses well below the electroweak scale, the two leading partial decay widths can be expressed as~\cite{Bauer:2017ris,Bauer:2018uxu,Chang:2021myh}
\begin{align}
&\label{eq:Gammall} \Gamma_{a\rightarrow\ell^{+}\ell^{-}} = \frac{(c^A_{\ell})^2 m^2_{\ell} m_a}{8\pi\Lambda^2}\sqrt{1-\frac{4m^2_{\ell}}{m^2_a}} \,, \\ 
&\label{eq:Gammaaa} \Gamma_{a\rightarrow\gamma\gamma} = \frac{m^3_a}{64\pi}\left(\frac{\alpha_{\text{em}}}{\pi}\frac{c^A_{\ell}}{\Lambda}\lvert 1 - {\cal F} (\frac{m^2_a}{4m^2_{\ell}})\rvert \right)^2  \,,
\end{align}
where the loop functions are given by ${\cal F} (z > 1) = \frac{1}{z}\text{arctan}^2\left(\frac{1}{\sqrt{1/z -1}}\right)$ and ${\cal F} (z < 1) = \frac{1}{z}\text{arcsin}^2\left(\sqrt{z}\right)$, respectively. 

The above discussion is general to all three generation of leptons. Until now, multiple works have been proposed to search for leptophilic ALPs~\cite{Kirpichnikov:2020lws,Bollig:2020xdr,Croon:2020lrf,Han:2020dwo,Chang:2021myh,Carenza:2021pcm,Cheung:2021mol,Caputo:2021rux,Bertuzzo:2022fcm,Cheung:2022umw,Lucente:2022esm,Altmannshofer:2022ckw,Lu:2022zbe,Calibbi:2022izs,Lu:2023ryd,Buonocore:2023kna,Calibbi:2024rcm,Budhraja:2024diy,Wang:2024zky,Jiang:2024cqj,Li:2025yzb,Li:2025beu,Yue:2025ksr,Huang:2025rmy,Ema:2025bww, Fiorillo:2025sln}. Among them, tauphilic ALPs have not drawn significant attention compared to the other two light flavors because of a more suppressed production. However, it is noted that a tauphilic ALP can also provide a rich phenomenon and complement studies in other facilities~\cite{Ema:2025bww,Belle-II:2022heu}. Some recent studies report that precise measurements of $\tau$ lepton dipole moments may assist the tauphilic ALP detection~\cite{Hoferichter:2025ijh, Hoferichter:2025zjp}. In this work, we focus on another avenue and investigate the potential of tauphilic ALPs in a $\tau$-factory, namely the STCF. Due to the presence of $aVV^{(\prime)}$ terms arising from the chiral anomaly and the novel four-point, $aW\tau\nu$, terms which depend on the choices of the coefficients $c^V_{\tau}$, $c^A_{\tau}$, and $c_{\nu}$, we identify two specific scenarios related to the electroweak symmetry structure~\cite{Altmannshofer:2022ckw,Jiang:2024cqj}: 
\begin{align} 
& \text{Electroweak Violating}~ (\text{EWV}) : c^V_{\tau} = c_{\nu} = 0, c^A_{\tau} \neq 0, \notag \\ 
& \text{Electroweak Preserving}~ (\text{EWP}) : c_{\nu} = 0, c^V_{\tau} = c^A_{\tau} \neq 0\,. 
\label{eq:EWV_EWP}
\end{align} 
In the EWV scenario, the lepton current in Eq.~(\ref{eq:Jint}) is purely axial-vector, while in the EWP scenario, it corresponds to right-handed coupling. Additionally, in the EWV scenario, all terms in Eq.~(\ref{eq:int}) are involved. In contrast, for the EWP scenario, only the $a\tau^+\tau^-$, $a\gamma\gamma$, $aZZ$, and $aZ\gamma$ interactions in Eq.~(\ref{eq:int}) remain~\footnote{It is crucial to clarify that the parameterization of ALP interaction in Ref.~\cite{Ema:2025bww} differs from ours in a surface term. Their benchmark is somewhat equivalent to our EWP case.}. 

\begin{figure*}[ht!]
\includegraphics[width=0.8\linewidth]{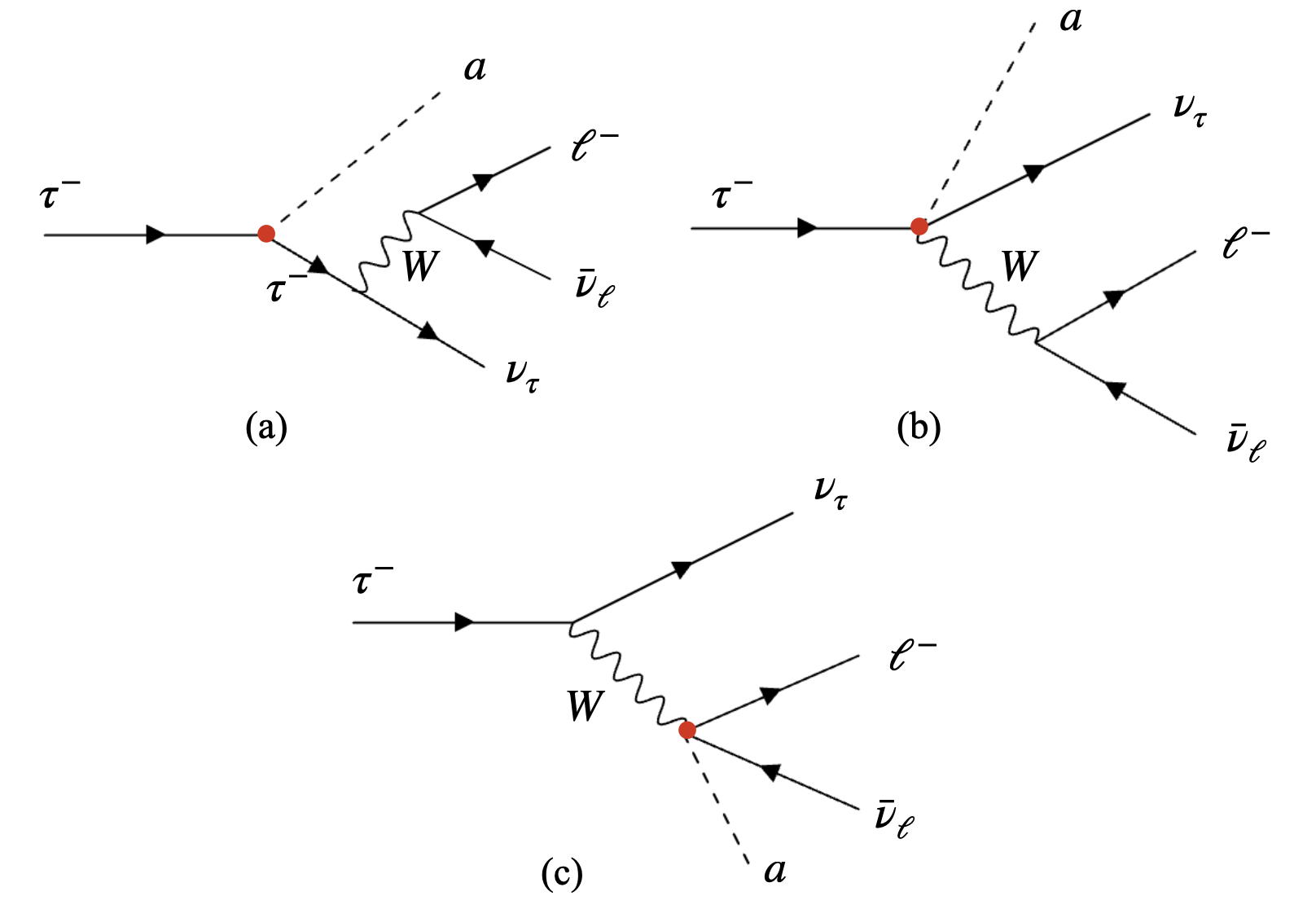}
\caption{The Feynman diagrams of ALP productions via leptonic $\tau$ decays. The red dot represents the vertex involving an ALP.}
\label{fig:feyn}
\end{figure*} 

\begin{figure*}[ht!]
\includegraphics[width=0.8\linewidth]{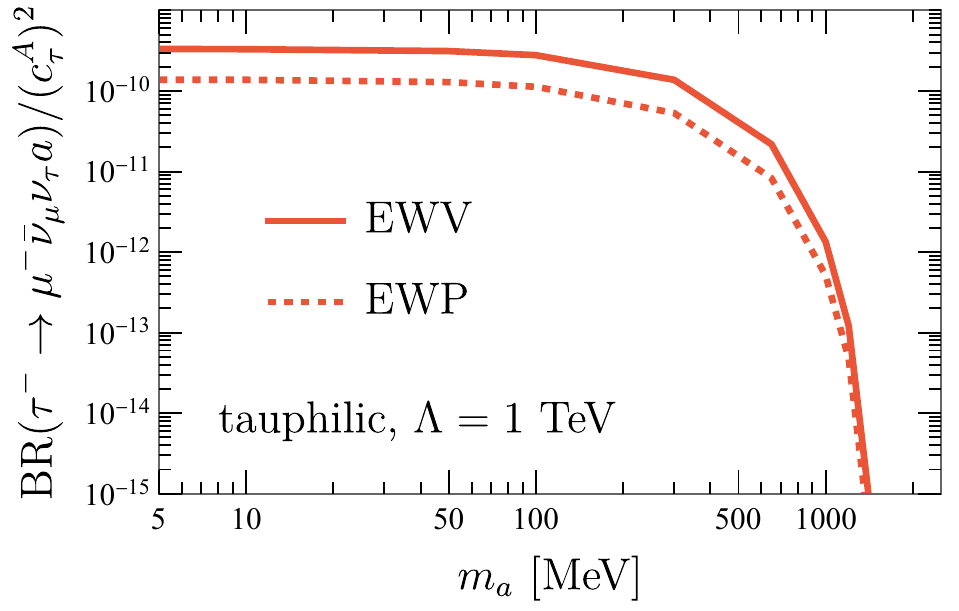}
\caption{The branching ratios for ALP productions via leptonic $\tau$ decays. The EWP and EWV scenarios are shown for the $\mu$-mode decay as dashed and solid curves, respectively. The corresponding $e$-mode decay yields nearly identical results (within $\sim 5\%$), and is omitted for clarity. 
}
\label{fig:braxiongeneration}
\end{figure*}

In a $\tau$ factory, ALPs are produced mainly via exotic decays of $\tau$-leptons. In this work, for simplicity, we focus purely on leptonic decays, $i.e.$ $\tau^- \to \ell^- \bar\nu_l \nu_\tau a$, with $\ell$ equal to $e$ or $\mu$. The related Feynman diagrams are presented in FIG.~\ref{fig:feyn}. In the EWP scenario, only FIG.~\ref{fig:feyn}(a) contributes to the ALP production. Different from light lepton flavors, the ALP radiation tends to be relevant thanks to a far heavier $\tau$-lepton mass. When it turns to the EWV scenario, both FIG.~\ref{fig:feyn}(a) and (b) are involved. Therefore, an enhancement in ALP production arises from the 4-point interaction. However, the story here significantly differs from ALP production from exotic meson decays, as discussed in~\cite{Altmannshofer:2022ckw,Jiang:2024cqj}. Notably, here both two Feynman diagrams take place at their tree level. As a result, regarding the EWP and EWV scenarios, no order difference in branching ratios should be expected. This is validated numerically. Concretely, we simulate $\tau$ decays associated with ALPs using \texttt{MadGraph5\_aMC@NLO~2.6.7}~\cite{Alwall:2014hca} and show the branching ratios in FIG.~\ref{fig:braxiongeneration}. 

In terms of the EWP scenario, our results show a high consistency with those of Ref.~\cite{Ema:2025bww}. It is interesting that the branching ratio of the $\mu$-mode tends to be slightly lower than that of the $e$-mode (around $5\%$), which may arise mainly from the charged lepton mass difference. In addition, we find that the branching ratios in the EWV scenario are always about 2 to 3 times larger than those in the EWP scenario, because of the existence of the additional four-point interaction. Finally, the branching ratio of $\tau$ decays is remarkably suppressed to $\mathcal O(10^{-10})$ when $c^A_\tau=1$. They tend to be stable, especially when the ALP mass falls below $\mathcal{O}(100)$ MeV. However, as the ALP increases to 1000 MeV, they are crucially suppressed by orders. Therefore, the high-intensity STCF experiment is ideal for probing such tau exotic decays.

\begin{figure}[h]
\centering 
\includegraphics[width=8 cm]{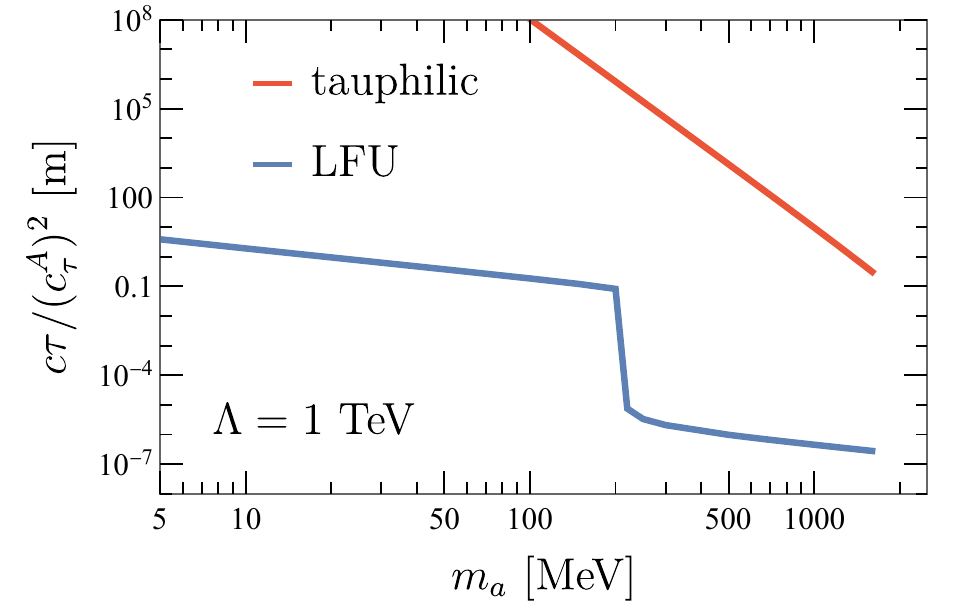}
\includegraphics[width=8 cm]{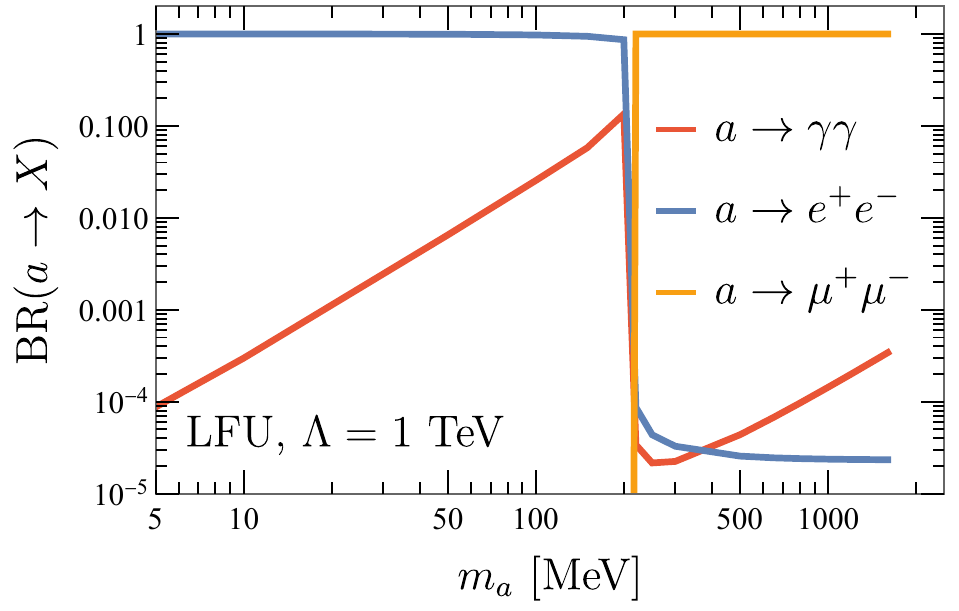}
\caption{\textbf{Left:} The lifetime of leptophilic ALP, with red and blue lines representing tauphilic and LFU benchmarks, respectively. \textbf{Right:} When LFU is assumed, the branching ratios of the ALP decays are presented in red, blue and orange lines, showing $a\to \gamma\gamma$, $a\to e^+e^-$ and $a\to \mu^+\mu^-$, respectively.}
\label{fig: alp_decay}
\end{figure}

In the tauphilic benchmark, the ALP is allowed only to decay into a pair of photons, as formulated in Eq.~(\ref{eq:Gammaaa}). Its lifetime is shown in the left panel of FIG.~\ref{fig: alp_decay}. Clearly, it is accordingly extended to a rather large scale, even exceeding $10^8$ meters when $m_a < 100$ MeV and $c^A_\tau=1$. 
A sufficiently long lifetime makes the ALP largely decay far away from the production vertex and consequently enriches phenomenology at far detectors of the STCF.

Alternatively, another attractive scenario is to consider letpon flavor universality (LFU). As one of the most important hypotheses, despite mass differences, all three leptons are assumed to be universally coupled to other particles in the SM. To validate the SM precisely and search for possible new physics, various studies would be found proposing a precise test of LFU at future colliders~\cite{Li:2020bvr, Ho:2022ipo, Deng:2025hio}. Here, in our work, we presume that the LFU holds in the ALP sector. Concretely, we set $c^{A,V}_\tau=c^{A,V}_\mu=c^{A,V}_e$. Both the production and decay of the ALP would be influenced. First and foremost, we can safely neglect ALP radiation from light leptons, since the interaction is severely suppressed by the lepton mass. As a result, only one more Feynman diagram should be included, which has been shown as FIG.~\ref{fig:feyn}(c). Interestingly, at an energy scale much lower than the electroweak one, FIG.~\ref{fig:feyn}(b) and (c) return opposite signs to cancel with each other at the amplitude level. Conclusively, the four-point interaction hardly contributes to the $\tau$ decay when LFU is assumed. The branching ratios of leptonic $\tau$ decays in the EWV scenario are thus found approximately equal to those in the EWP scenario. For the LFU case analyzed in the following section, we only focus on the EWV scenario, since the choice of $c_\ell^V$ has negligible impact, as argued above.

The ALP decay in the LFU scenario is more complicated. The lifetime and branching ratios of ALP are provided in FIG.~\ref{fig: alp_decay}. When the ALP mass is heavier than $2m_\mu$, it is allowed to decay into diphoton, dielectron and dimuon. When it is lighter than $2m_\mu$ but heavier than $2m_e$, only diphoton and dielectron channels are switched on. It is reported that dimuon and dielectron channels dominate when the ALP is heavier and lighter than $2m_\mu$, respectively. Nevertheless, the diphoton channel is irrelevant in this scenario. Consequently, the ALP lifetime is tremendously shortened below 100 m when $c^A_\tau=1$. A step-like structure is expected near the kinematic threshold $2m_\mu$, manifesting in both the lifetime and branching ratios. 
The shorter lifetime makes detection at the main detector more promising. Due to the ALP decay branching ratios shown in FIG.~\ref{fig: alp_decay}, our analysis can be largely simplified. For a given ALP mass, only one decay channel should be emphasized in STCF analyzes at a time.
More details on collider analysis would be found in the next section.

\section{Search strategies for leptophilic ALPs at the STCF and it far detectors}\label{sec:tas} 
\subsection{Tauphilic ALPs}\label{sec:taualp}
As discussed in the previous section, a tauphilic ALP can play a crucial role in exotic decays of the $\tau$-lepton. The abundant production of $\tau$-lepton at the STCF acts as a perfect facility for detecting such signals, showing relevant projections of the ALP parameter space. In this study, two benchmarks are taken into account. On the one hand, we consider the central energy $\sqrt s=3.56$ GeV. As evaluated in Refs.~\cite{Achasov:2023gey,Pich:2024qob}, for a one-year run, more than $10^8$ $\tau^+\tau^-$ pairs can be collected. Here, the beam energy falls below the $D^0\bar D^0$ threshold, and the beam energy spread is expected to be well controlled below 1 MeV~\cite{Achasov:2023gey}. Based on these two facts, the STCF keeps free from heavy-quark backgrounds to create a super neat environment to evaluate exotic $\tau$ decays. On the other hand, a higher central energy $\sqrt s=4.2$ GeV is also investigated. Approximately $3.6\times 10^{9}$ $\tau^+\tau^-$ pairs per year will be available at the STCF~\cite{Achasov:2023gey}. The search profits from larger statistics, but meanwhile suffers from possible backgrounds from $c$-hadrons. To simplify the analysis, in this work, we purely investigate leptonic $\tau$ decays and believe that, in the leptonic sector, the background would be well controlled to a negligible level when long-lived ALPs are considered. For both benchmarks, we simulate $e^+e^-\to \tau^+ \tau^-$ events at the STCF where one tau lepton undergoes the exotic decay, $\tau^-\to a\ell^-\bar{\nu_{\ell}}\nu_{\tau}$ or $\tau^+\to a\ell^+\bar{\nu_{\tau}}\nu_{\ell}$, using \texttt{MadGraph5\_aMC@NLO~2.6.7}~\cite{Alwall:2014hca}\footnote{Here a minimal transverse momentum cut of $p_t (\tau) > 0.1$ GeV is applied at the parton level.}. 
The tauphilic ALPs can only decay into diphoton in both EWP and EWV scenarios. 

\begin{figure}[h]
\centering 
\includegraphics[width=8 cm]{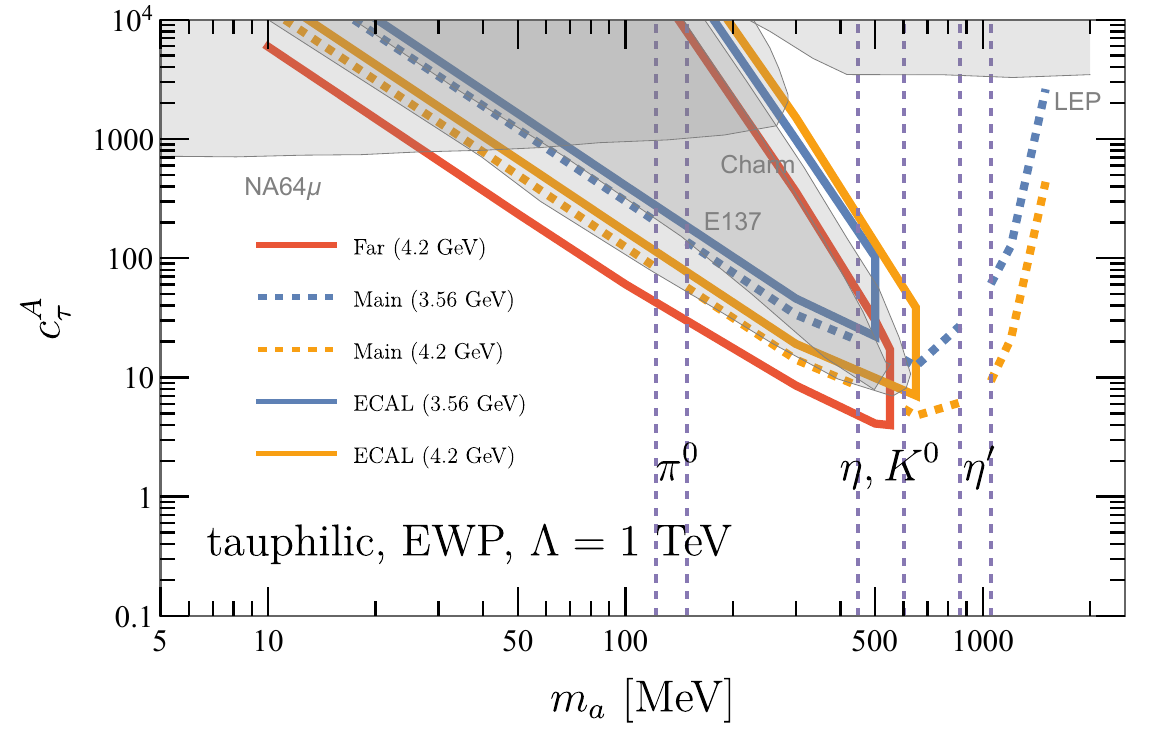}
\includegraphics[width=8 cm]{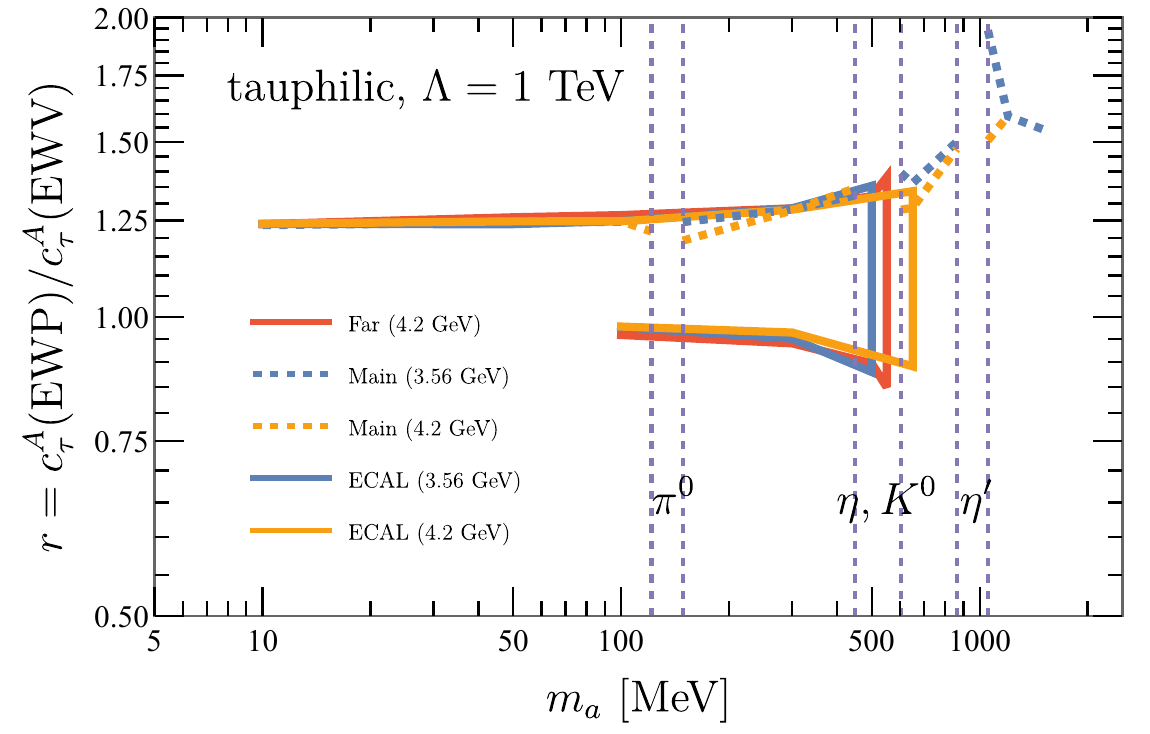}
    \caption{\textbf{Left:} In the EWP scenario, the projections at $95\%$ confidence level are achieved at the main and far detectors. The red curve stands for far detection. The blue and orange curves represent projections for main detector with $\sqrt s=3.56$ GeV and $\sqrt s=4.2$ GeV, respectively. Dashed and solid curves correspond to the main detection outside and inside the ECAL, respectively. Existing bounds from other facilities are also shown for comparison, including NA64~\cite{NA64:2024klw}, LEP~\cite{Jaeckel:2015jla}, E137~\cite{Bjorken:1988as, Eberhart:2025lyu}, and CHARM~\cite{CHARM:1985anb}. \textbf{Right: }Comparison between the EWP and EWV scenarios for main and far detections. }
\label{fig: farprompt}
\end{figure}

From the detector perspective, we explore the potential of detection in both the main and far detectors with ten-year-run data. We initiate with the far detection. Currently, proposals for far detectors at the STCF are still missing from the market. Hence, we simply take one possible setup for illustration in this work. The far detector is proposed to have a cylinder structure, with its front surface $10$ m away from the interacting point (IP), a radius of reaching $20$ m, and a length of up to $30$ m. Its orientation follows along the $z$-axis in the lab frame~\footnote{We have also checked the set-up along a transverse axis and comparable results have been found.}. The decay volume is about one-tenth of MATHUSLA~\cite{MATHUSLA:2022sze} and approximately 3 times larger than the ANUBIS shaft configuration~\cite{Bauer:2019vqk}. The detector is supposed to measure both charged particles and photons with the tracking system and electromagnetic calorimeter (ECAL), respectively. To capture the tauphilic ALP signal, the ALP must decay inside the detector, whose probability is formulated as
\begin{align}\label{eq:prob}
P &= e^{-\frac{l_1}{\beta\gamma\tau}}-e^{-\frac{l_2}{\beta\gamma\tau}}~,
\end{align}
where $\beta$, $\gamma$, and $\tau$ correspond to ALP velocity, Lorentz boost factor, and ALP lifetime, respectively. In addition, $l_1$ and $l_2$ denote the distances traveled by the ALP to enter and escape the detector, respectively. For far detection, only a small range of the solid angle is allowed, so that the performance is significantly affected by statistics. Consequently, only the $\sqrt s=4.2$ GeV benchmark is evaluated in a far detector. The projections reaching 95\% confidence level are obtained based on the background-free hypothesis and are shown in red solid lines of FIG.~\ref{fig: farprompt}.

Next, it is time to focus on the detection in the main detector. In both EWP and EWV scenarios, the ALP is forced to decay into a pair of photons, and these decay products are targeted at the main detector. Here are two possibilities. If the ALP travels and decays after entering the ECAL, we search for trackless diphoton signals to reconstruct the resonance. The photons from the SM processes are substantially suppressed, as they are likely to convert into $e^+ e^-$ pairs before reaching the ECAL. The probability of ALP decay can be parameterized with Eq.~(\ref{eq:prob}), by requiring $l_1=1$ m and $l_2=1.3$ m, respectively~\cite{Achasov:2023gey}. However, if the ALP decays between the IP and the ECAL, the vertex reconstruction would be challenging. Under this circumstance, we may only require $l_1=0$ and $l_2=1$ m, and assume that an invariant mass window cut can significantly suppress SM backgrounds. In both two cases, we require that every photon and the ALP candidate should have $p>100$ MeV, where the photon detection efficiency reaches 0.8~\cite{Achasov:2023gey}. To avoid the background of SM meson decays to diphoton, the regions in the vicinity of the $\pi^0$, $\eta$, $\eta^{\prime}$ and $K^0_L$ masses ($\pm 10\%$) have all been discarded. Eventually, the background-free hypothesis has been optimistically adopted, and the constraints at $95\%$ confidence level are shown in blue and orange lines of FIG.~\ref{fig: farprompt}, where the dashed and solid curves stand for projections obtained before and after arriving at the ECAL, respectively.

Here are some comments on our results. First and foremost, as shown in FIG.~\ref{fig: farprompt}, in the tauphilic scenario, neither the main nor a far one constrains the coupling for very light ALP masses. This is due to the extremely long ALP lifetime, which prevents decays within the detector volume. The far detector provides limited complementarity to existing experiments such as E137 and CHARM, with sensitivity comparable to proto-DUNE~\cite{Ema:2025bww}.
When it turns to the main detector, the projections obtained with the ECAL largely overlap with the existing ones, such as E137 and Charm, while they slightly extend the probe in the regime with $m_a\sim \mathcal O(400-500)$ MeV and $c_\tau^A\sim 100$. In contrast, the detection before the ECAL is more promising. It is particularly interesting to focus on the region where $m_a\sim 1000$ MeV and $c_\tau^A\sim \mathcal O(10^2-10^3)$. This region has not yet been probed, but can be fully explored at the STCF. Moreover, we find that the search at $\sqrt s = 4.2$ GeV takes advantage of large statistics and therefore provides a stronger bound compared to the one at the threshold. However, in this analysis, we simply consider leptonic decay and assume backgrounds from heavy quarks being irrelevant. This assumption may not hold when hadronic decays are taken into account. In such a case, the threshold scenario will show its advantage in a clean environment. We leave a sophisticated study on the hadronic $\tau$ decays associated with an ALP for future study.

In addition, the EWV scenario exhibits a larger contour than the EWP case, as clearly seen in the right panel of FIG.~\ref{fig: farprompt}. Here, $r$ is defined as the ratio of $c_\tau^A$ between the EWP and EWV scenarios at 95\% confidence level. Particularly, the detection in both the far detector and ECAL shows two branches for $r$ values. The part below $r = 1$ corresponds to the upper limit part presented in the left panel. In summary, all detection channels show a $(15–75)\%$ enhancement in the EWV scenario to constrain $c_\tau^A$, attributable to increased ALP production from the four-point interaction. In contrast to exotic meson decays~\cite{Altmannshofer:2022ckw,Jiang:2024cqj}, however, the difference between the two scenarios considered here is much smaller.

\subsection{ALPs with LFU}\label{sec:alplfu} 
\begin{figure}[h]
\centering 
\includegraphics[width=12 cm]{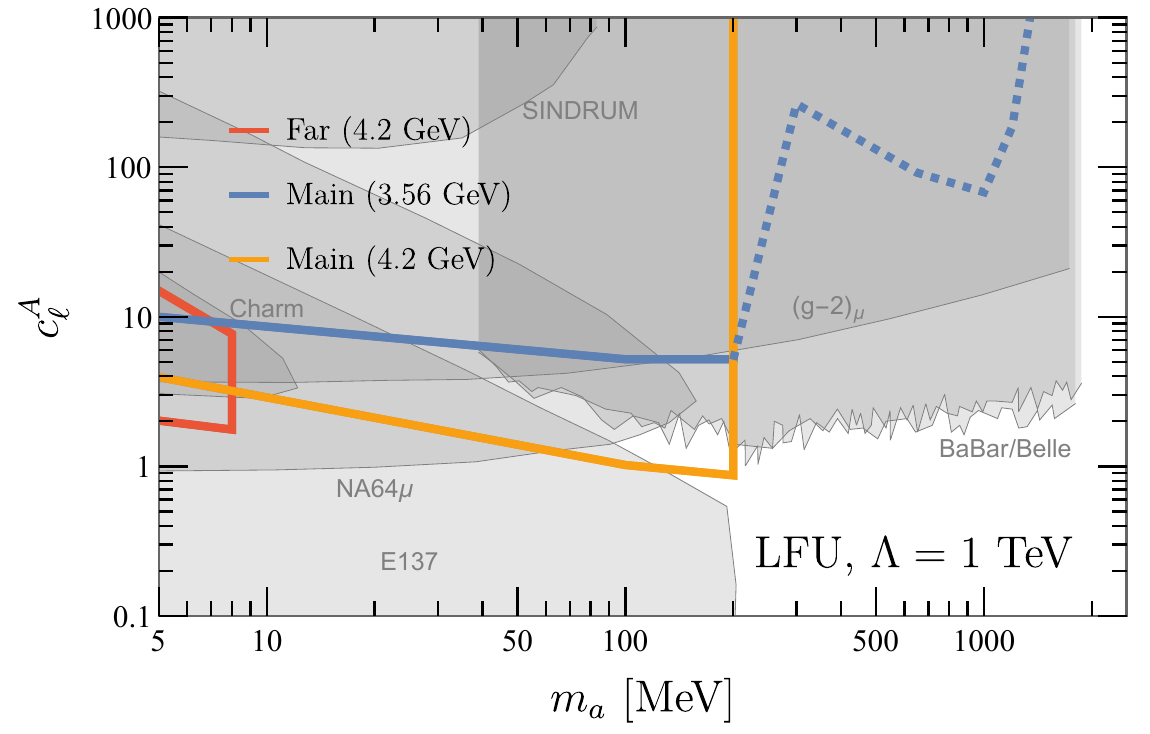}
\caption{The projections at $95\%$ confidence level for ALPs with LFU. The far detection, main detection at $\sqrt s = 3.56$ GeV and main detection at $\sqrt s = 4.2$ GeV are shown in red, blue and orange curves, respectively. The blue dashed curve corresponds to the projections achieved with $l_1=0$ at $\sqrt s=3.56$ GeV. For comparison, the limits achieved at other experiments are presented as shaded regions, including NA64~\cite{NA64:2024klw}, SINDRUM~\cite{SINDRUM:1986klz}, Charm~\cite{CHARM:1985anb}, E137~\cite{Bjorken:1988as, Eberhart:2025lyu}, BaBar~\cite{BaBar:2020jma}, Belle~\cite{Belle:2022gbl} and the latest muon $(g-2)_\mu$ measurement~\cite{Muong-2:2025xyk} with the SM prediction~\cite{Aliberti:2025beg}.}
\label{fig: farprompt_lfu}
\end{figure}

In the LFU scenario, light leptons dominate ALP decays. Consequently, as shown in FIG.~\ref{fig: alp_decay}, the ALP lifetime is substantially reduced, and the $a\to\gamma\gamma$ channel becomes negligible. The dominant decays instead proceed to dielectron or dimuon pairs, depending on the ALP mass. The final state contains three charged leptons. We require $p (e) > 100$ MeV, but $p (\mu) > 400$ MeV suggested by the muon detector~\cite{Achasov:2023gey}. For mixed-flavor final states ($e^+e^-\mu^-$ or $\mu^+\mu^-e^-$), the lepton pair of ALP decays is readily identified. We require its invariant mass lies within $(0.9, 1.1)m_a$. For same-flavor lepton triplets, we select the opposite-charge pair with invariant mass closest to $m_a$, requiring it to fall within the $10\%$ mass window, while other combinations lie outside. For both cases, we require the ALP candidate to have $p(a)>100$ MeV. Following Refs.~\cite{Achasov:2023gey, Bertholet:2025lcr}, we assume the tau trigger and muon detection efficiency to reach 0.9 and 0.8, respectively. In order to suppress prompt SM backgrounds like $\tau\to 3\ell+2\nu$, when $m_a < 2m_\mu$, we further require a displaced decay vertex, by setting $l_1=10^{-4}$ m and $l_2=1$ m. However, when $m_a > 2m_\mu$, the ALP lifetime shrinks considerably. In this case, the displaced vertex search with $l_1=10^{-4}$ m fails. We consider only $\sqrt s = 3.56$ GeV, which heavy-quark backgrounds are absent, and optimistically set $l_1=0$ m for illustration. We derive $95\%$ confidence level constraints under the background-free hypothesis, shown in FIG.~\ref{fig: farprompt_lfu}. For comparison, we include projections from NA64~\cite{NA64:2024klw}, SINDRUM~\cite{SINDRUM:1986klz}, CHARM~\cite{CHARM:1985anb}, E137~\cite{Bjorken:1988as, Eberhart:2025lyu}, BaBar~\cite{BaBar:2020jma}, Belle~\cite{Belle:2022gbl}, and the latest muon $(g-2)_\mu$ measurement~\cite{Muong-2:2025xyk}. As explained in Sec.~\ref{sec:alpproduction}, the amplitude from FIG.~\ref{fig:feyn}(b) cancels that of FIG.~\ref{fig:feyn}(c) at low energies. Consequently, the projections for both EWP and EWV scenarios are almost identical. Therefore, we display only the EWV scenario in FIG.~\ref{fig: farprompt_lfu} for illustration. 

Here are some comments. First and foremost, in the LFU scenario, ALP decays to light leptons are enabled, significantly shortening the ALP lifetime compared to the tauphilic ALP scenario. Consequently, ALPs with masses as low as several MeV may decay within the detector volume, making them promising targets for the STCF. This lifetime reduction also affects far detection: ALPs decay more promptly, substantially shrinking the viable parameter space for far detection signals at the STCF. As shown in FIG.~\ref{fig: farprompt_lfu}, the far detection sensitivity is generally comparable to that of CHARM but extends to slightly smaller values of \(c_\ell^A\). 
The far detection is most sensitive for ALPs with \(\mathcal{O}(1)\) MeV masses and \(\mathcal{O}(1)\) couplings. However, the main detection at the \(\tau^+\tau^-\) threshold energy cannot probe this region effectively due to the limited \(\tau^+\tau^-\) production rates. With ten-year-run data, threshold measurements largely overlap existing constraints and provide minimal new information. It is interesting to point out that a spike has been found around the dimuon mass threshold in FIG.~\ref{fig: farprompt_lfu}. The dashed part corresponds to the detection above the threshold, where the main decay channel switches from dielectron to dimuon. There are several issues here. Firstly, a harder muon momentum cut has been applied, causing a vital suppression of the signal efficiency. When increasing the mass of ALP, a higher signal efficiency becomes possible due to the larger kinematic phase space. Secondly, the dimuon decay channel shortens the ALP lifetime and makes it difficult to produce a sufficiently displaced decay vertex. However, for this ALP mass region, the Barbar and Belle experiments can relevantly constrain the ALP couplings.


The situation improves markedly at \(\sqrt{s} = 4.2\) GeV for detection at the main detector. The higher center-of-mass energy yields greater statistics, leading to stronger exclusion bounds. As shown in FIG.~\ref{fig: farprompt_lfu}, the projected constraints at this energy cover the regions accessible to both far detection partly and searches at the $\tau^+\tau^-$ threshold, and extend significantly to include ALP masses from $100$ MeV to $200$ MeV with couplings near unity. Similarly to other future experiments such as FASER II, SHiP, and SHiNESS~\cite{Ema:2025bww, Jiang:2024cqj, Wang:2024zky}, limited sensitivity is achieved at the STCF when the ALP exceeds 200 MeV, where the ALPs mostly decay within $l_1 < 10^{-4}$ m, since the lifetime is suppressed due to switching on the decay channel $a\to \mu^+\mu^-$. A systematic study of relevant QCD backgrounds becomes necessary and may potentially assist exploration in the currently unprobed parameter space around \(m_a \sim 500\) MeV and \(0.1 < c_\ell^A < 1\). However, it goes beyond the scope of this work and we leave it for future study.

\section{Conclusions}\label{sec:conclusion}

In this work, we fully investigate the potential for detecting leptophilic axion-like particles (ALPs) via exotic leptonic $\tau$ decays at the future $\tau$-factory, namely Super-Tau-Charm-Facility (STCF). Two scenarios are taken into full consideration. One is related to tauphilic ALPs and the other presumes ALPs with lepton flavor universality (LFU). Multiple decay channels of the ALP are well evaluated, including both $a\to \gamma\gamma$ and $a\to \ell^+\ell^-$ ($\ell = e\text{ or }\mu$). A comprehensive assessment of the reach at the STCF is provided. Several collider set-ups are considered. On the one hand, main detection is explored at either $\tau^+\tau^-$ threshold ($\sqrt s = 3.56$ GeV) or a higher energy that maximizes the production rate ($\sqrt s = 4.2$ GeV). Furthermore, a far detector is proposed to complement the analysis at the main detector. 

Regarding tauphilic ALPs, we find only $a\to\gamma\gamma$ is allowed and such a channel remarkably extends the ALP lifetime to prevent decays taking place inside the detector volume. Therefore, the obtained projections tend to be narrow, just as shown in FIG.~\ref{fig: farprompt}. Notably, far detection complements those in the main detector. The ECAL can act as an independent detector to search for trackless photon signals and meanwhile, the main detector can be used to detect ALP decays before they reach the ECAL by applying a specific invariant mass window cut. In general, all these searches fail when $m_a$ drops to 10 MeV. However, new opportunities have been found. The ECAL extends the search to $m_a\sim \mathcal O(400-500)$ MeV and $c_\tau^A\sim \mathcal O(10^2)$. On the other hand, the main detector manages to probe the yet untouched region with $m_a$ around 1 GeV and $c_\tau^A\sim \mathcal O(10^2)$. Additionally, we evaluated both the electroweak violating (EWV) and preserving (EWP) scenarios. 
In contrast to meson decays, where EWV and EWP scenarios exhibit significantly different sensitivities~\cite{Jiang:2024cqj,Altmannshofer:2022ckw}, the EWV scenario here yields only an
$\mathcal O(20\%)$ improvement over EWP. This stems from the absence of loop suppression in ALP production of the EWP scenario (See FIG.~\ref{fig:feyn}(a)) and only the presence of an additional four-point interaction in the EWV scenario (See FIG.~\ref{fig:feyn}(b)).

In the LFU case, all the interactions involving light leptons have been switched on. Firstly, we point out that the four-point interaction no longer plays an important role in ALP production via exotic $\tau$ decays, since the amplitude from FIG.~\ref{fig:feyn}(b) cancels with that from FIG.~\ref{fig:feyn}(c) at low energies. Therefore, EWP and EWV tend to be identical and do not deserve separate discussions. In addition, due to the involvement of light leptons, the decay $a\to \ell^+\ell^-$ becomes the dominant channel. This significantly reduces the ALP lifetime, enhancing the prospect of detection at the main detector. The projections are acquired and shown in FIG.~\ref{fig: farprompt_lfu}. Explicitly, the contour of the far detection shrinks and is shown to be comparable to Charm~\cite{CHARM:1985anb}. The main detection at the $\tau^+\tau^-$ threshold is limited by the $\tau$ production rate and does not provide new constraints, whereas a higher energy scenario takes advantage of larger statistics and is competitive in probes of the unexplored parameter space, where $m_a\sim 0.2$ GeV and $c_\ell^A\sim 1$. 

\section*{Acknowledgments}
We thank Zeren Simon Wang for helpful discussions and Martin Hoferichter for insightful comments. 
XHJ and CTL are supported in part by the National Natural Science Foundation of China (NNSFC) under grant No.~12342502, No.~12335005, No.~12575118, and the Special funds for postdoctoral overseas recruitment, Ministry of Education of China. 
The authors gratefully acknowledge the valuable discussions and insights provided by the members of the China Collaboration of Precision Testing and New Physics.

\end{document}